\documentclass[]{ima}
\usepackage{amsmath}
\usepackage{amsfonts}
\usepackage{amssymb}
\usepackage{epsfig}
\usepackage{epstopdf}
\usepackage{color, url}

\newtheorem{ex}[theorem]{Example}

\newcommand{\CC}{\mathbb C}
\newcommand{\RR}{\mathbb R}

\newcommand{\ZZ}{\mathbb Z}

 \newcommand{\T}{\mathcal{T}}
  \newcommand{\cT}{\mathcal{T}}

    \newcommand{\bL}{\mathbb{L}}
        \newcommand{\mbA}{\mathbf{A}}
    \newcommand{\ind}{\mathrm{index}}
    \newcommand{\conv}{\mathrm{conv}}

\begin{document}

\markboth{BERND STURMFELS and JOSEPHINE YU}{TROPICAL IMPLICITIZATION}

\title{TROPICAL IMPLICITIZATION \\ AND MIXED FIBER POLYTOPES}

\author{BERND STURMFELS\thanks{University of California, Berkeley, CA 94720, bernd@math.berkeley.edu}
 \and JOSEPHINE YU\thanks{Massachusetts Institute of Technology, Cambridge, MA 02139,
 jyu@math.mit.edu}}
 \maketitle

\date{\today}

\begin{abstract}                   
The software TrIm offers implementations of tropical implicitization and tropical elimination,
as developed by Tevelev and the authors.
Given a polynomial map with generic coefficients, TrIm
computes the tropical variety of the image.  When the image is a hypersurface, 
the output is the Newton polytope of the defining polynomial.
TrIm can thus be used to compute 
 mixed fiber polytopes, including
secondary polytopes.
\end{abstract} 


{\AMSMOS 14Q10, 52B20, 52B55, 65D18
\endAMSMOS}

 \section{Introduction}
 
Implicitization is the problem of transforming a given parametric representation of an algebraic variety into its implicit representation as the zero set of polynomials.  Most algorithms for elimination and implicitization are  based on multivariate resultants or Gr\"obner bases, but current implementations of these methods are often  too slow. When the variety is a hypersurface,
the coefficients of the implicit equation can also be computed 
by way of numerical linear algebra \cite{CGKW, EK},
provided the Newton polytope of that implicit equation can be predicted a priori.

The problem of predicting the Newton polytope was recently solved
independently by three sets of authors, namely, 
by Emiris,  Konaxis and Palios \cite{EKP}, Esterov and Khovanskii \cite{KE},
and in our joint papers with Tevelev \cite{ST, STY}.
A main conclusion of these papers can be summarized as follows:
{\em The Newton polytope of the implicit equation is a mixed fiber polytope}.

The first objective of the present article is to explain this conclusion 
and to present the software package {\tt TrIm} for computing such
mixed fiber polytopes. The name of our program stands for {\em Tropical Implicitization},
and it underlines our view that the prediction of Newton polytopes is best 
understood within the larger context of tropical algebraic geometry.
The general theory of tropical elimination developed in \cite{ST}
unifies earlier results on discriminants  \cite{DFS} and on
generic polynomial maps whose images can have any codimension \cite{STY}.
The second objective of this article is to explain
the main results of tropical elimination theory
and their implementation in {\tt TrIm}.
Numerous hands-on examples will illustrate the use of the software.
At various places we give precise pointers to \cite{EKP} 
and \cite{KE}, so as to highlight similarities and differences
among the different approaches to the subject.

Our presentation is organized as follows.
In Section 2 we start out with a quick guide to {\tt TrIm}
by showing some simple computations. In Section 3 we
explain mixed fiber polytopes. That exposition is self-contained
and may be of independent interest to combinatorialists.
In Section 4 we discuss the computation of mixed fiber polytopes
in the context of elimination theory, and in Section 5 we show how
the tropical implicitization problem is solved in {\tt TrIm}.
Theorem 5.1 expresses the Newton polytope of the
implicit equation as a mixed fiber polytope.
In Section 6 we present  results in tropical
geometry on which the development of {\tt TrIm} is based,
and we explain various details concerning our
algorithms and their implementation.

\section{How to use {\tt TrIm}}
The first step is to download {\tt TrIm} from the following 
website which contains information for installation in {\tt Linux}:
\smallskip
\begin{center}
  \url{http://math.mit.edu/~jyu/TrIm}
\end{center}

\smallskip

\noindent
{\tt TrIm} is a collection of {C++} programs 
which are glued together and 
integrated with the external software {\tt polymake} \cite{GJ} using {\tt perl} scripts.  
The language
{\tt perl} was chosen for ease of interfacing between various programs.

The fundamental problem in tropical implicitization is
to compute the Newton polytope of a hypersurface
which is parametrized by Laurent polynomials with
sufficiently generic coefficients.
As an example we consider the following three Laurent polynomials in
two unknowns $x$ and $y$ with sufficiently generic coefficients
$\alpha_1,\alpha_2,\alpha_3, \beta_1, \beta_2,\beta_3,\gamma_1,\gamma_2,\gamma_3$:
\begin{eqnarray*}
u \quad & = & \quad \alpha_1 \cdot \frac{1}{x^2y^2} \,+\, \alpha_2 \cdot x \,+\, \alpha_3 \cdot xy \\
v \quad & = & \quad \beta_1 \cdot x^2 \,+\, \beta_2 \cdot y \,+\, \beta_3 \cdot \frac{1}{x}  \\
w \quad & = & \quad \gamma_1 \cdot y^2 \,+\, \gamma_2 \cdot \frac{1}{xy} \,+\, \gamma_3 \cdot \frac{1}{y} .
\end{eqnarray*}
We seek the unique (up to scaling) irreducible polynomial
$F(u,v,w)$ which vanishes on the image of the corresponding
 morphism $ (\CC^*)^2 \rightarrow \CC^3$.
Using our software {\tt TrIm}, the Newton polytope of
the polynomial $F(u,v,w)$ can be computed as follows.
We first create a file {\tt input} with the contents

 \small \begin{verbatim}

[x,y]
[x^(-2)*y^(-2) + x + x*y, 
 x^2 + y + x^(-1), 
 y^2 + x^(-1)*y^(-1) + y^(-1)]

\end{verbatim} 
 \normalsize

\smallskip

\noindent
Here the coefficients are suppressed: they are tacitly 
assumed to be generic.
We next run a {\tt perl} script using the command {\tt ./TrIm.prl input}.
The output produced by this program call is quite long. It includes the lines

\small \begin{verbatim}

VERTICES
1  9 2 0 
1  0 9 2 
1  0 9 0 
1  0 0 9 
1  6 6 0 
1  2 2 8 
1  0 6 6 
1  2 8 2 
1  6 0 6 
1  0 0 0 
1  9 0 0 
1  2 0 9 
1  8 2 2 

\end{verbatim} \normalsize
Ignoring the initial {\tt 1}, this list consists of $13$ lattice points in $\RR^3$,
and these are precisely the vertices of the Newton polytope of $F(u,v,w)$.
The above output format is compatible with the polyhedral software {\tt Polymake} \cite{GJ}.
We find that the Newton polytope has
$10$ facets, $21$ edges, and $13$ vertices.
Further down in the output, {\tt TrIm} prints a list of all lattice points
in the Newton polytope, and it ends by telling us the number of lattice points:

\small \begin{verbatim}

N_LATTICE_POINTS
383

\end{verbatim} \normalsize
Each of the $383$ lattice points $(i,j,k)$ represents a monomial
$u^i v^j w^k$ which might occur with non-zero
coefficient in the expansion of $F(u,v,w)$.
Hence, to recover the coefficients of $F(u,v,w)$ we must solve
a linear system of $382$ equations with $383$ unknowns. Interestingly, in this example,
$39$ of the $383$ monomials always have coefficient zero in $F(u,v,w)$.
Even when $\alpha_1,\ldots,\gamma_3$ are completely generic,
the number of monomials in $F(u,v,w)$ is only $344$.

The command {\tt ./Trim.prl} implements a certain algorithm,
to be described in the next sections, whose input consists
of $n$ lattice polytopes in $\RR^{n-1}$ and whose output
consists of one lattice polytope in $\RR^n$. In our example,
with $n=3$, the input consists of three triangles and the
output consisted of a three-dimensional polytope.
These are depicted in Figure 1.

\begin{figure}

\begin{tabular}{cc}
 \includegraphics[scale=0.6]{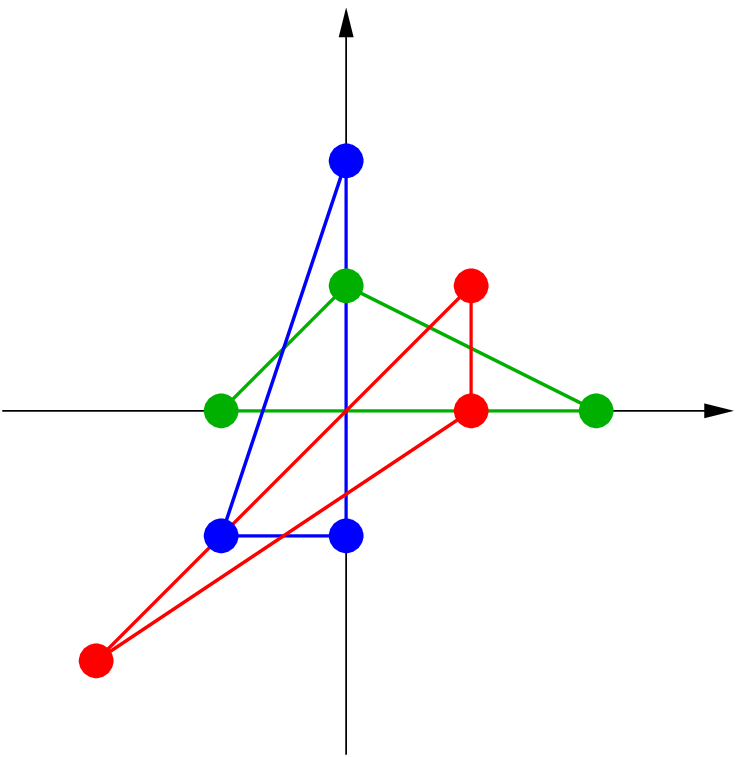} &
 \hskip-1.20cm \includegraphics[scale=0.5]{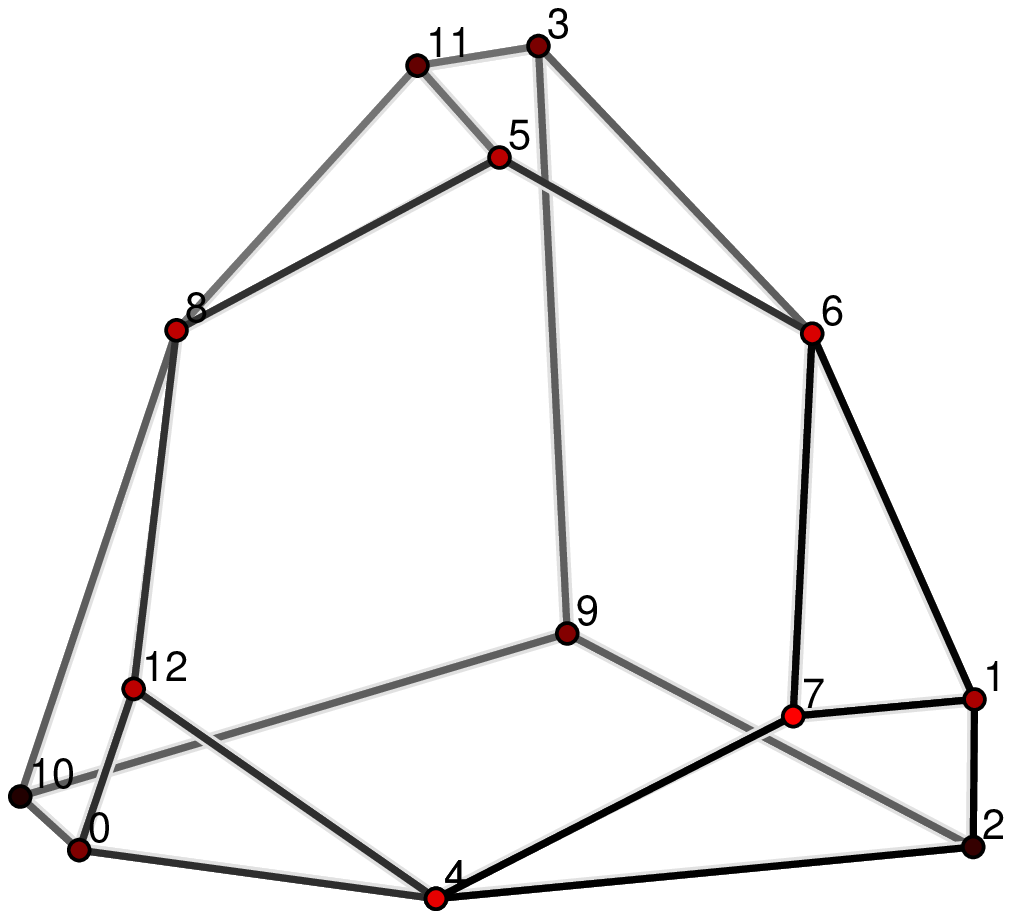} \\
Input & Output
\end{tabular}

\caption{Tropical implicitization constructs
the three-dimensional Newton polytope of a parametrized surface
from the three Newton polygons of the given parametrization}

\end{figure}

The program also works in higher dimensions but
the running time quickly increases. For instance,
consider the hypersurface in $\CC^4$ represented by 
the following  four Laurent polynomials in
$x,y,z$, written in {\tt TrIm} format:

\small \begin{verbatim}

[x,y,z]
[x*y + z + 1,
 x*z + y + 1,
 y*z + x + 1,
 x^3 + y^5 + z^7]

\end{verbatim} \normalsize
It takes ${\tt TrIm}$ a few moments to inform us that the Newton polytope of  this
hypersurface has $40$ vertices and contains precisely $5026$ lattice points.
The  $f$-vector of this four-dimensional
polytope equals $(40, 111, 103, 32)$.

\smallskip

\begin{remark} \rm The examples above may serve as
illustrations for the results in the papers  \cite{EKP} and \cite{KE}.
Emiris, Konaxis and Palios \cite{EKP} place the emphasis
on computational complexity, they present a 
precise formula for plane parametric curves, and they
allow for the map to given by rational functions.
Esterov and Khovanskii develop a general theory
of polyhedral elimination, which parallels the tropical
approach in \cite{ST}, and which includes implicitization
as a very special case. A formula for the
leading coefficients of the implicit equation is given in
\cite[\S 4]{EKP}. This formula is currently not implemented
in {\tt TrIm} but it could be added in a future version.
\end{remark}

\smallskip

What distinguishes {\tt TrIm} from the approaches in \cite{EKP}
and \cite{KE} is the command {\tt TrCI} which 
computes the tropical variety of a generic complete intersection.
The relevant mathematics will be reviewed in Section 6.
This command is one of the ingredients in the implementation
of tropical implicitization. The input again consists of
$m$ Laurent polynomials in $n$ variables
whose coefficients are tacitly assumed to be generic,
or, equivalently, of $m$ lattice polytopes in $n$-space.
Here it is assumed that $m \leq n$. If equality holds then
the program simply computes the mixed volume
of the given polytopes. As an example, delete
the last line from the previous input file:

\small \begin{verbatim}

[x,y,z]
[x*y + z + 1,
 x*z + y + 1,
 y*z + x + 1]
 
\end{verbatim} \normalsize
The command  {\tt ./TrCI.prl input} computes
the mixed volume of the three given 
lattice polytopes in $\RR^3$.
Here the given polytopes are triangles.
The last line in the output shows that their
mixed volume equals five.

Now repeat the experiment with the input file {\tt input} as follows:

\small \begin{verbatim}
[x,y,z]
[x*y + z + 1,  x*z + y + 1]

\end{verbatim} \normalsize
The output is a one-dimensional
tropical variety given by five rays in $\RR^3$:

\small \begin{verbatim}

DIM
1

RAYS
 0 -1 -1
 0  0  1
 1  0  0
 0  1  0
-1  1  1

MAXIMAL_CONES
0
1
2
3
4

MULTIPLICITIES
2
1
1
1
1

\end{verbatim} \normalsize
Note that the first ray, here indexed by {\tt 0}, has multiplicity two.
This scaling ensures that the sum of the five {\tt RAYS} equals the zero vector $(0,0,0)$.

For a more interesting example, let us tropicalize the
complete intersection of two generic hypersurfaces in $\CC^5$.
We prepare {\tt input} as follows:

\small \begin{verbatim}

[a,b,c,d,e]
[a*b + b*c + c*d + d*e + a*e + 1,
 a*b*c + b*c*d + c*d*e + d*e*a + e*a*b]

\end{verbatim} \normalsize
When applied to these two polynomials in five unknowns,
the command   {\tt ./TrCI.prl input} produces
a three-dimensional fan in $\RR^5$.
This fan has $26$ rays and it has $60$
 maximal cones. Each maximal cone is the cone over a 
 triangle or a quadrangle, and it has multiplicity one.
 The rays are

\smallskip

 \small \begin{verbatim}

RAYS
-1  1  0  0  1
-1  1  1 -1  3
 0  1  0  0  1
-1  3 -1  1  1
\end{verbatim} \normalsize
$ \,\,\, \cdots \quad  \cdots \quad \cdots $

\smallskip

\noindent
The rays are labeled $0,1,\ldots,25$, in the order in which they
were printed. The maximal (three-dimensional) cones
appear output in the format

\small \begin{verbatim}

MAXIMAL_CONES
0 1 2 3
0 1 7 10
0 1 12
0 3 4 7
0 3 12
0 7 12
\end{verbatim} \normalsize
$ \, \cdots \,\,\,  \cdots  $

It is instructive to compute the tropical intersection
of two generic hypersurfaces with the same support.
For example, consider the input file

\small \begin{verbatim}

[x,y,z]
[1 + x + y + z + x*y + x*z + y*z + x*y*z,
 1 + x + y + z + x*y + x*z + y*z + x*y*z]

\end{verbatim} \normalsize
As before, the reader should imagine
that the coefficients are generic rational numbers instead of one's.
The tropical complete intersection
determined  by these two equations consists of the six rays
normal to the six facets of the given three-dimensional cube.
The same output would be produced by Jensen's
software {\tt GFan} \cite{BJSST, gfan}, which computes arbitrary tropical varieties,
 provided we input the two equations
with generic coefficients.

\section{Mixed fiber polytopes}

We now describe the construction of mixed fiber polytopes.
These generalize ordinary fiber polytopes \cite{BS}, and hence they generalize
secondary polytopes \cite{GKZ}. The existence of mixed fiber polytopes 
was predicted by McDonald \cite{McD}  and
Michiels and Cools \cite{MC} in the context of polynomial systems solving. They
were first constructed by McMullen \cite{McM}, and later
independently by Esterov and Khovanskii \cite{KE}.

The presentation in this section is written entirely
in the language of combinatorial geometry, and it should be of
independent interest to some of the readers of Ziegler's text book \cite{Zie}.
There are no polynomials or varieties in this section, 
neither classical nor tropical.
The connection to elimination and tropical geometry
will be explained in subsequent sections.

Consider a linear map $\, \pi : \RR^p \rightarrow \RR^q$
and a $p$-dimensional polytope $P \subset \RR^p$ whose image $Q = \pi(P)$ is a
$q$-dimensional polytope in $\RR^q$.
If $x$ is any point in the interior of $Q$ then
its fiber $\,\pi^{-1}(x) \cap P\,$ is a polytope of
dimension $p-q$. The {\em fiber polytope}
is defined as the Minkowski integral 
\begin{equation}
\label{minkint}
 \Sigma_\pi(P) \quad = \quad 
\int_Q (\pi^{-1}(x) \cap P) \,dx. 
\end{equation}
It was shown in \cite{BS} that this
integral defines a
 polytope of dimension $p-q$.
The fiber polytope $\Sigma_\pi(P)$ lies in an
affine subspace of $\RR^p$ which is
a parallel translate of $\, {\rm kernel}(\pi)$.
Billera and Sturmfels \cite{BS} used the notation $\Sigma(P,Q)$
for the fiber polytope, and they
showed that its faces are in bijection with the
coherent polyhedral subdivisions of $Q$ which 
are induced from the boundary of $P$.
 We here prefer the notation $\Sigma_\pi(P)$
 over the notation $\Sigma(P,Q)$, so as to
highlight the dependence on $\pi$ 
for fixed $P$ and varying $\pi$.

\smallskip

\begin{ex} \rm \label{3cube}
Let $p = 3$ and take $P$ to be the standard $3$-cube
$$ P \,\,\, = \,\,\, {\rm conv}\bigl\{ 
(000), (001), (010), (011),
(100), (101), (110), (111) \bigr\}.
$$
We also set $q=1$ and we fix the linear map
$$ \pi :  \RR^3 \rightarrow \RR^1, \, (u,v,w) \mapsto  u+2v+3w .$$
Then $Q = \pi(P)$ is the line segment $[0,6]$.
For $0 < x < 6$, each fiber $\pi^{-1}(x) \cap P$ is either
a triangle, a quadrangle or a pentagon. Since the fibers
have a fixed normal fan over each open segment $(i,i+1)$, 
we find
$$
\Sigma_\pi(P) \quad = \quad
\sum_{i=0}^5 \int_{i}^{i+1} \! (\pi^{-1}(x) \cap P) \,dx
\quad = \quad
\sum_{i=0}^5 \bigl(\pi^{-1}(i+ \frac{1}{2}) \cap P \bigr).
$$
Hence the fiber polygon is really just the
 Minkowski sum of two triangles,
two quadrangles and two pentagon,
and this turns out to be a hexagon:
$$ \Sigma_\pi(P) \,\, = \,\, 
{\rm conv} \bigl\{    (1, 10, 5),
    (1, 4, 9),    (5, 2, 9),
   (11, 2, 7),   (11, 8, 3),       (7, 10, 3) \bigr\}. $$
In the next section we shall demonstrate how {\tt TrIm}
can be used to compute fiber polytopes. The output produced
will be the planar hexagon which is gotten
from the coordinates above by applying the linear map
$(u,v,w) \mapsto (w-3, v+w-9)$. Hence {\tt TrIm}
produces the following coordinatization:
\begin{equation}
\label{FiberHexagon}
\Sigma_\pi(P) \,\,\, = \,\, \,
{\rm conv} \bigl\{
 (2, 6), \,
 (6, 4), \,
 (6, 2), \,
 (4, 0), \,
 (0, 2), \,
 (0, 4) \,
 \bigr\}.
\end{equation}
It is no  big news to polytope aficionados that the
 fiber polygon of the $3$-cube is a
hexagon. Indeed, by  \cite[Example 9.8]{Zie},
 the fiber polytope obtained by projecting
the $p$-dimensional cube onto a line
is the {\em permutohedron}
of dimension $p-1$. For $p=3$
the vertices of the hexagon $\Sigma_\pi(P)$ correspond to
the six monotone edge paths on
the $3$-cube from $(000)$ to $(111)$. \qed
\end{ex}
\smallskip

As a special case of the construction of fiber polytopes we get
the secondary polytopes.
Suppose that $P$ is a polytope with $n$ vertices
in $\RR^p$ and let $\Delta$ denote the standard $(n-1)$-simplex
in $\RR^n$. There exists a linear map
$\,\rho : \RR^n \rightarrow \RR^p$ such that
$\rho(\Delta) = P$, and this linear map is unique
if we prescribe a bijection from
the vertices of $\Delta$ onto the vertices of $P$.
The polytope $\Sigma_\rho(\Delta)$ is called
the {\em secondary polytope} of $P$; see   \cite[Definition 9.9]{Zie}.
Secondary polytopes were first introduced in an algebraic context by
 Gel'fand, Kapranov and Zelevinsky \cite{GKZ}.
For example, if we take $P$ to be the $3$-dimensional cube as above,
then the simplex $\Delta$ is $7$-dimensional,
and the secondary polytope $\Sigma_\rho(\Delta)$ is
a $4$-dimensional polytope with $74$ vertices.
These vertices are in bijection with the $74$ triangulations of the $3$-cube.

A detailed introduction to triangulations and a range of 
methods for computing secondary polytopes
can be found in the forthcoming book \cite{DRS}.
We note that the computation of fiber polytopes
can in principle be reduced to the computation of secondary polytopes,
by means of the formula
\begin{equation}
\label{compose}  \Sigma_\pi( \rho(\Delta)) \,\,\, = \,\,\,
\rho(\Sigma_{\pi \circ \rho}(\Delta)).
\end{equation}
Here $\pi \circ \rho$ is the composition of the following two linear maps of polytopes:
$$ \Delta \stackrel{\rho}{\longrightarrow} P \stackrel{\pi}{\longrightarrow} Q $$
The formula (\ref{compose}) appears in \cite[Lemma 2.3]{BS} and in
\cite[Exercise 9.6]{Zie}.
The algorithm of Emiris {\it et al.} \cite[\S 4]{EKP} for computing Newton polytopes
of specialized resultants is based on a variant of (\ref{compose}). 
Neither our software {\tt TrIm} nor the
Esterov-Khovanskii construction \cite{KE} uses 
the formula (\ref{compose}).

We now come to the main point of this section, namely,
the construction of {\em mixed fiber polytopes}. This is primarily due to
McMullen \cite{McM}, but was rediscovered 
in the context of elimination theory by
Khovanskii and  Esterov \cite[\S 3]{KE}.
We fix a linear map $\pi : \RR^p \rightarrow \RR^q$ as above,
but we now consider a collection of $c$ polytopes
$P_1,\ldots,P_c$ in $\RR^p$. We consider the Minkowski sum
$\,P_\lambda=\lambda_1 P_1 + \cdots + \lambda_c P_c\,$
where $\, \lambda = (\lambda_1,\ldots,\lambda_c)$
is a parameter vector of unspecified positive real numbers. 
We shall assume that  $P_\lambda$ is of full dimension $p$,
but we do allow its summands
$P_i$ to be lower-dimensional. The image of $P_\lambda$
under the map $\pi$ is the $q$-dimensional polytope
$$ \pi(P_\lambda) \,\,\, = \,\,\, \lambda_1 \cdot \pi(P_1) + \cdots + \lambda_c \cdot \pi(P_c). $$
The following result concerns the fiber polytope
from $\,P_\lambda\,$ onto $\,\pi(P_\lambda)$.

\smallskip

\begin{theorem}[\cite{McM, KE}]
The fiber polytope
 $\Sigma_\pi(P_\lambda)$ 
 depends polynomially on the parameter vector $\lambda$. 
This polynomial is homogeneous of degree $q+1$.
 Moreover, there exist unique polytopes $\,M_{i_1 i_2 \cdots i_c}\,$ such that
\begin{equation}
\label{decompo}
\Sigma_\pi (\,\lambda_1 P_1 + \cdots + \lambda_c P_c \,)
\,\,\,\,\, = \sum_{i_1 + \cdots + i_c = q+1} \!\!\!\!\!\!\!
\lambda_1^{i_1}
\lambda_2^{i_2} \cdots \lambda_c^{i_c} \cdot M_{i_1 i_2 \cdots i_c}.
\end{equation}
\end{theorem}

To appreciate this theorem, it helps to begin with the case $c=1$.
That corresponds to scaling the polytopes $P$ and $Q$ above by the same factor $\lambda$.
This results in the Minkowski  integral (\ref{minkint})  being scaled by the factor $\lambda^{q+1}$.
More generally, the coefficients of the pure powers $\lambda_j^{q+1}$
in the expansion  (\ref{decompo}) are precisely the fiber polytopes of the individual $P_j$, that is,
$$ M_{0,\ldots,0,q+1,0,\ldots,0} \,\, = \,\, \Sigma_\pi(P_j) . $$
On the other extreme, we may consider $\,i_1 = i_2 = \cdots = i_c = 1$, which
 is the term of interest for elimination theory. Of course, 
if all  $i_j$'s are equal to one then the number $c$ of polytopes $P_j$ is 
one more than the dimension $q$ of the image of $\pi$. We now assume that this
holds, i.e., we assume that $\, c = q+1 $. We define the
 {\em mixed fiber polytope} to be the coefficient of the monomial
 $\lambda_1\lambda_2 \cdots \lambda_c $ in the formula (\ref{decompo}).
The mixed fiber polytope is denoted
\begin{equation}
\label{MFP}\Sigma_\pi(P_1,P_2, \ldots,P_c) \quad := \quad M_{11\cdots 1}.
 \end{equation}
The smallest non-trivial case arises when $p=3$, $c=2$ and $q=1$,
where we are projecting two polytopes $P_1$ and $P_2$ in $\RR^3$.
Their mixed fiber polytope with respect to a linear form
$\,\pi : \RR^3 \rightarrow \RR^1 \,$ is the coefficient of $\,\lambda_1 \lambda_2\,$ in
$$ \Sigma_\pi( \lambda_1 P_1  + \lambda_2 P_2) \quad = \quad
\lambda_1^2 \cdot \Sigma_\pi(P_1)\, \,+\,\,
\lambda_1 \lambda_2 \cdot \Sigma_\pi(P_1,P_2)  \,\,+\,\,
\lambda_2^2 \cdot \Sigma_\pi(P_2) .$$
The following is \cite[Example 4.10]{ST}.
It will be revisited in Example \ref{ccurve}.

\smallskip 

\begin{ex} \rm
\label{twotetrahedra}
Consider the following two tetrahedra in  three-space:
$$\, P_1 =  {\rm conv}\{ 0, 3 e_1, 3 e_2, 3 e_3\} \, \hbox{ and } \,
 P_2 =  {\rm conv}\{ 0, -2 e_1, -2 e_2, -2 e_3\}. $$
  Their Minkowski sum $P_1 + P_2$ has
   $12$ vertices, $24$ edges and $14$ facets.
  If we take  $ \pi : \RR^3 \rightarrow \RR^1$ 
  to be the linear form $\, (u,v,w) \mapsto u-2v+w \,$
  then the fiber polytope 
 $\, \Sigma_\pi(  P_1 + P_2) \,= \, M_{20} \,+\,  M_{11}  \, +\,  M_{02} \,$
 is a polygon with ten vertices. Its summands
  $M_{20} = \Sigma_\pi(P_1) $ and  $M_{02} = \Sigma_\pi(P_2) $ are quadrangles,
  while the mixed fiber polytope   $\,M_{11} = \Sigma_\pi(P_1,P_2)\,$ is
  a hexagon. \qed
 \end{ex}

\smallskip
 
 We remark that fiber polytopes
are special instances of mixed fiber polytopes.
Suppose that $P_1 = P_2 =  \cdots = P_c$ are
all equal to the same fixed polytope $P$ in $\RR^p$.
Then the fiber polytope $\Sigma_\pi (P_\lambda)$
 in (\ref{decompo}) equals 
$$ \Sigma( \lambda_1 P_1 + \cdots + \lambda_c P_c )
\quad = \quad (\lambda_1 + \cdots + \lambda_c)^c \cdot \Sigma_\pi(P) .$$
Hence the fiber polytope $\Sigma_\pi(P)$ is
the mixed fiber polytope $ \Sigma_\pi(P, \ldots,P) $
scaled by a  factor of $1/c !$.
Similarly, any of the
coefficients in the expansion (\ref{decompo})
can be expressed as mixed fiber polytopes.
Up to scaling, we have
$$\,M_{i_1 i_2 \cdots i_c} \quad = \quad\,\,
 \Sigma_\pi \bigl(
\underbrace{P_1,\ldots,P_1}_{i_1 \,{\rm times}},
\underbrace{P_2,\ldots,P_2}_{i_2 \,{\rm times}},\ldots,
\underbrace{P_c,\ldots,P_c}_{i_c \, {\rm times}} \bigr). $$
In the next section we shall explain how
mixed fiber polytopes,
and hence also fiber polytopes
and secondary polytopes, can be computed using {\tt TrIm}.

\section{Elimination}

Let $\,f_1,f_2,\ldots,f_c \in \CC[x_1^{\pm 1},x_2^{\pm 1},\ldots,x_p^{\pm 1}]\,$
be Laurent  polynomials whose Newton polytopes are $P_1,P_2,\ldots,P_c \subset \RR^p$,
and suppose that the coefficients of the $f_i$ are generic.
This means that 
$$ f_i(x) \quad = \quad \sum_{a \in P_i \cap \ZZ^n} c_{i,a} \cdot x_1^{a_1} x_2^{a_2}
\cdots x_p^{a_p}, $$
where the coefficients $c_{i,a}$ are assumed to be sufficiently generic
non-zero complex numbers. The corresponding variety
$$ X \quad = \quad \bigl\{
u \in (\CC^*)^p \,\,:\,\, f_1(u) = f_2(u) = \cdots = f_c(u) = 0 \bigr\} $$
is a complete intersection of codimension $c$ 
in the algebraic torus $(\CC^*)^p$.

We set $r = p-c+1$ and we
fix an integer  matrix ${\bf A} = (a_{ij})$ of format $r \times p$
where the rows of ${\bf A}$ are assumed to be linearly independent.
We also let
 $\pi : \RR^p \rightarrow \RR^{c-1}$ be any linear map
whose kernel equals the row space of ${\bf A}$.
The matrix ${\bf A}$ induces the following monomial map:
\begin{equation}
\label{monomap}
 \alpha : (\CC^*)^p \rightarrow (\CC^*)^{p-c+1}\,,\,\,\,
(x_1,\ldots,x_p) \mapsto \bigl( 
\prod_{j=1}^p x_j^{a_{1j}},\ldots,
\prod_{j=1}^p x_j^{a_{rj}} \bigr). 
\end{equation}
Let $Y$ be the closure in $(\CC^*)^{p-c+1}$
of the image $\alpha(X)$. Then $Y$ is a hypersurface,
and we are interested in its Newton polytope. By this we mean
the Newton polytope of the irreducible equation of that hypersurface.

\smallskip

\begin{theorem}[Khovanskii and Esterov \cite{KE}]
\label{tropelim}
The Newton polytope of $\,Y\,$ is affinely isomorphic to
the mixed fiber polytope $\Sigma_\pi(P_1,\ldots,P_c)$.
\end{theorem}

\smallskip

A proof of this result using tropical geometry is given in \cite{ST}.
The computation of the hypersurface $Y$ from the defining
equations $f_1,\ldots,f_c$ of $X$ is a key problem of 
elimination theory. Theorem \ref{tropelim} offers a
tropical solution to this problem. It predicts the Newton
polytope of $Y$. This information is useful for
symbolic-numeric software. Knowing the Newton polytopes reduces
computing the equation of $Y$ to linear algebra.

The numerical mathematics  of this linear algebra problem is 
interesting and challenging, as seen in \cite{CGKW} and confirmed
by the experiments reported in \cite[\S 5.2]{STY}.
We hope that our software {\tt TrIm} will eventually
be integrated with software exact linear algebra or numerical
linear algebra (e.g. {\tt LAPack}).
Such a combination would have the potential of becoming a useful tool
for practitioners of non-linear computational geometry.

\smallskip

In what follows, we demonstrate how {\tt TrIm} 
computes  the Newton polytope of $Y$ and hence
the mixed fiber polytope $\Sigma_\pi(P_1,\ldots,P_c)$.
The input consists of
the polytopes $P_1,\ldots,P_c$ and the matrix ${\bf A}$.
The map $\pi$ is tacitly understood as the map
from $\RR^p$ onto the cokernel of the transpose of ${\bf A}$.

\smallskip

\begin{ex} \rm \label{ccurve}
Let $p=3,c=2$ and consider \cite[Example 1.3]{ST}.
Here the variety $X$ is the curve in $(\CC^*)^3$ defined by the 
two Laurent polynomials
$$ f_1 \,=\,  \alpha_1 x_1^3 +  \alpha_2 x_2^3 +  \alpha_3 x_3^3 +  \alpha_4 
\quad \hbox{and} \quad
f_2 \,=\,  \beta_1 x_1^{-2} +  \beta_2 x_2^{-2} +  \beta_3 x_3^{-2} +  \beta_4 . $$
We seek to compute the Newton polygon 
of the image curve $Y$ in $(\CC^*)^2$ where  ${\bf A} = 
\begin{pmatrix} 
1 &  1 & 1 \\ 
0 & 1 & 2 
\end{pmatrix}$. The curve is written on a file {\tt input}  as follows:

\small \begin{verbatim}

[x1,x2,x3]
[x1^3 + x2^3 + x3^3 + 1, x1^(-2)+x2^(-2)+x3^(-2)+1]

\end{verbatim} \normalsize
We also prepare a second input file {\tt A.matrix} as follows:

\small \begin{verbatim}

LINEAR_MAP
1 1 1
0 1 2

\end{verbatim} \normalsize
We now execute the following two commands in {\tt TrIm}:

\small \begin{verbatim}

./TrCI.prl input > fan
./project.prl fan A.matrix 

\end{verbatim} \normalsize
The output we obtain is the Newton polygon of the curve $Y$:

\small \begin{verbatim}

VERTICES
1 36  0 
1  0 36 
1 30 12 
1 18 12 
1  6 24 
1 18 24 

\end{verbatim} \normalsize
This hexagon coincides with the hexagon in
 \cite[Examples 1.3 and 4.10]{ST}. It is isomorphic to
 the mixed fiber polytope $\Sigma_\pi(P_1,P_2)$
 in Example \ref{twotetrahedra}. \qed
\end{ex}

\smallskip

We may use {\tt TrIm} to compute arbitrary fiber polytopes.
For example, to carry out the computation of Example  \ref{3cube}, we
prepare {\tt input} as

\small \begin{verbatim}

[x,y,z]
[1 + x + y + z + x*y + x*z + y*z + x*y*z,
 1 + x + y + z + x*y + x*z + y*z + x*y*z]

\end{verbatim} \normalsize
and {\tt A.matrix} as

\small \begin{verbatim}

LINEAR_MAP
1  1 -1
2 -1  0

\end{verbatim} \normalsize
The two commands above now produce the hexagon in (\ref{FiberHexagon}).
Our next example shows how to compute secondary polytopes using {\tt TrIm}.

\smallskip

\begin{ex} \rm
Following \cite[Example 9.11]{Zie}, we consider the hexagon
with vertices $(0,0), (1,1), (2,4), (3,9), (4,16), (5,25)$.
This hexagon is represented in {\tt TrIm} by the following 
file ${\tt A.matrix}$. The rows of this matrix span the linear
relations among the five non-zero vertices of the hexagon:

\small \begin{verbatim}

LINEAR_MAP
 3   -3  1  0  0
 8   -6  0  1  0
15  -10  0  0  1

\end{verbatim} \normalsize
On the file {\tt input} we take
three copies of the standard $5$-simplex:

\small \begin{verbatim}

[a,b,c,d,e]
[a+b+c+d+e+1,  a+b+c+d+e+1, a+b+c+d+e+1]
 \end{verbatim} \normalsize
 Running our two commands, we obtain
 a  $3$-dimensional polytope with $14$ vertices,
 $21$ edges  and $9$ facets. That polytope is the {\em associahedron} \cite{Zie}. \qed
\end{ex}

\smallskip

We close this section with another application of tropical elimination.

\smallskip

\begin{ex} \rm
For two subvarieties $X_1$ and $X_2$ of $(\CC^*)^n$
we define their {\em coordinate-wise product}
$ X_1 \star X_2 $ to be the closure of the set of all points
$ (u_1 v_1,\ldots,u_n v_n)$ where $(u_1,\ldots,u_n) \in X_1$ 
and $(v_1,\ldots,v_n) \in X_2$. The expected dimension of $X_1 \star X_2$
is the sum of the dimensions of $X_1$ and $X_2$, so we can expect
$X_1 \star X_2$ to be a hypersurface when ${\rm dim}(X_1) + {\rm dim}(X_2) = n-1$.
Assuming that $X_1$ and $X_2$ are generic complete intersections
then the Newton polytope of that hypersurface can be computed
using {\tt TrIm} as follows. Let  $p = 2n$ and define $X$ as the
direct product $X_1 \times X_2 $. Then $X_1 \star X_2$ is the image of $X$
under the monomial map
$$ \alpha : (\CC^*)^{2n} \rightarrow (\CC^*)^n ,\,  (u_1,\ldots,u_n, v_1,\ldots,v_n) \mapsto 
(u_1 v_1,\ldots,u_n v_n). $$
Here is an example where $X_1$ and $X_2$ are curves in
three-dimensional space $(n=3)$.
The two input curves are specified on the file {\tt input} as follows:

\small \begin{verbatim}

[u1,u2,u3, v1,v2,v3]
[u1 + u2 + u3 + 1,
 u1*u2 + u1*u3 + u2*u3 + u1 + u2 + u3,
 v1*v2 + v1*v3 + v2*v3 + v1 + v2 + v3 + 1,
 v1*v2*v3 + v1*v2 + v1*v3 + v2*v3 + v1 + v2 + v3]

\end{verbatim} \normalsize
The multiplication map $\alpha : (\CC^*)^3 \times
(\CC^*)^3 \rightarrow (\CC^*)^3$ is specified on {\tt A.matrix}:

\small \begin{verbatim}

LINEAR_MAP
1 0 0 1 0 0 
0 1 0 0 1 0 
0 0 1 0 0 1

\end{verbatim} \normalsize
The image of $X_1 \times X_2$ under the map $\alpha$ is the
surface $X_1 \star X_2 $.
We find that the Newton polytope of this surface has
 ten vertices and seven facets:

\small \begin{verbatim}

VERTICES
1 8 4 0 
1 0 8 4 
1 0 8 0 
1 0 0 8 
1 4 8 0 
1 4 0 8 
1 0 4 8 
1 8 0 0 
1 0 0 0 
1 8 0 4 

FACETS
128 -16   0   0
  0   0   0  32
128   0 -16   0
  0  32   0   0
  0   0  64   0
128   0   0 -16
192 -16 -16 -16

\end{verbatim} \normalsize
\end{ex}

\section{Implicitization}

Implicitization is a special case of elimination.
Suppose we are given $n$ Laurent polynomials
$g_1,\ldots,g_{n}$ in $\CC[t_1^{\pm 1}, \ldots, t_{n-1}^{\pm 1}]$
which have Newton polytopes $Q_1,\ldots,Q_n \subset \RR^{n-1}$
and whose coefficients are generic complex numbers. These data defines the morphism
\begin{equation}
\label{morphismG}
 \, g \, : \, (\CC^*)^{n-1} \rightarrow (\CC^*)^n\,,\,
t \,\mapsto\, \bigl(\, g_1(t),\ldots,g_{n-1}(t) \bigr) .
\end{equation}
Under mild hypotheses, the closure of the image of $g$ is
a hypersurface $Y$ in $(\CC^*)^n$. Our problem is to 
compute the Newton polytope of this hypersurface.
A first example of how this is done in {\tt TrIm} was shown
in the beginning of Section 2, and more examples will be featured in this section.

The problem of implicitization is reduced to the elimination computation in the previous 
section as follows. We introduce $n$ new variables $y_1,\ldots,y_n$
and we consider the following $n$ auxiliary  Laurent polynomials:
\begin{equation}
\label{auxf}
 f_1(x) \, = \, g_1(t) \,-\, y_1 \,, \,\,\,
     f_2 (x) \, = \, g_2(t) \,-\, y_2 \,, \,\,
      \ldots , \,
     f_n (x) \, = \, g_n(t) \, - \, y_n . 
     \end{equation}
Here we set $p = 2n-1$ and  $(x_1,\ldots,x_p) = (t_1,\ldots,t_{n-1},y_1,\ldots,y_n)$
so as to match the earlier notation. The subvariety of
$\,(\CC^*)^p \, = \, (\CC^*)^{n-1} \times (\CC^*)^n \,$ defined by
$f_1,\ldots,f_n$ is a generic complete intersection of codimension $n$, namely,
it is the graph of the map $g$. The image of $g$ is
obtained by projecting the
variety $\,\{f_1 = \cdots = f_n = 0\}\,$ onto the last $n$ coordinates. This 
projection is the monomial map $\alpha$ specified
by the $n \times p$-matrix $A~=~(\mathbf{0}~I)$ where $\mathbf{0}$ is the $n \times (n{-}1)$ matrix of zeroes and $I$ is the $n \times n$ identity matrix.

This shows that we can solve the implicitization problem by
doing the same calculation as in the previous section. Since
that calculation is a main application of {\tt TrIm}, we have
hard-wired it in the command {\tt ./TrIm.prl}. Here is an example
that illustrates the advantage of using tropical implicitization in
analyzing parametric surfaces of high degree in three-space.

\smallskip

\begin{ex} \rm
Consider the parametric surface specified by the {\tt input} 

\small \begin{verbatim}

[x,y]
[x^7*y^2 + x*y + x^2*y^7 + 1,
 x^8*y^8 + x^3*y^4 + x^4*y^3 + 1,
 x^6*y + x*y^6 + x^3*y^2 + x^2*y^3 + x + y]

\end{verbatim} \normalsize
Using the technique shown in Section 2, we learn in
a few seconds to learn that the irreducible
 equation of this surface has degree $90$.
The command {\tt ./TrIm.prl input} reveals that its
Newton polytope has six vertices

\small \begin{verbatim}

VERTICES
1  80  0  0
1   0 45  0
1   0  0 80
1   0 10 80
1   0  0  0
1  28  0 54

\end{verbatim} \normalsize
This polytope also has six facets, namely four triangles and two quadrangles.
The expected number of monomials in the implicit equation equals

\small \begin{verbatim}

N_LATTICE_POINTS
62778

\end{verbatim} \normalsize
At this point the user can make an informed choice as to
whether she wishes to attempt solving for the coefficients
using numerical linear algebra.  \qed
\end{ex}

\smallskip

Returning to our polyhedral discussion in Section 3, we next give a 
conceptual formula for the Newton polytope of the implicit equation
as a mixed fiber polytope. The given input is a list of $n$ lattice polytopes
$Q_1,Q_2,\ldots,Q_n$ in $\RR^{n-1}$. Taking the direct product of 
$\RR^{n-1}$ with the space $\RR^n$ with standard basis $\{e_1,e_2,\ldots,e_n\}$,
we consider the auxiliary polytopes
$$ \conv(Q_1 \cup \{e_1\}) \,,  \,\,
      \conv(Q_2 \cup \{e_2\}) \, , \,\, \ldots \,, \,\, \conv(Q_n \cup \{e_n\})
      \,\,\, \subset \,\,\, \RR^{n-1} \times \RR^n,$$
      where $\conv$ denotes the convex hull.
These are the Newton polytopes of the equations 
$f_1,f_2,\ldots,f_n$ in (\ref{auxf}).
We now define $\pi$ to be the projection onto the first $n-1$ coordinates
$$ \pi :  \RR^{n-1} \times \RR^n \rightarrow \RR^{n-1} \,, \,\,
(u_1,\ldots,u_{n-1}, v_1,v_2, \ldots,v_n) \,\mapsto \, (u_1,\ldots,u_{n-1}).$$
The following result is an immediate corollary to
Theorem \ref{tropelim}. We propose that it be named the
{\em Fundamental Theorem of Tropical Implicitization}.

\smallskip

\begin{theorem}
The Newton polytope of an irreducible hypersurface 
in $(\CC^*)^n$ which is
parametrically represented by generic Laurent polynomials with
given Newton polytopes
$Q_1,\ldots,Q_n$ equals the mixed fiber polytope
\begin{equation}
\label{mfpimp}
\Sigma_\pi \bigl( \conv(Q_1 \cup \{e_1\}) ,
     \conv( Q_2 \cup \{e_2\})  ,  \ldots ,  \conv(Q_n \cup \{e_n\}) \bigr).
 \end{equation}
\end{theorem}

This theorem is the geometric characterization of the Newton
polytope of the implicit equation, and it summarizes the
essence of the recent progress obtained by
Emiris, Konaxis and Palios \cite{EKP}, Esterov and Khovanskii \cite{KE}, 
and Sturmfels, Tevelev and Yu \cite{ST, STY}.
Our implementation of in {\tt TrIm} computes the
mixed fiber polytope (\ref{mfpimp}) for any given
$Q_1,Q_2,\ldots,Q_n$, and it suggests that
the Fundamental Theorem of Tropical Implicitization
will be a tool of considerable practical value for computational algebra. 

\smallskip

\begin{ex} \rm
We consider a threefold in $\CC^4$ which is parametrically
represented by four trivariate polynomials. On the file {\tt input} we write

\small \begin{verbatim}

[x,y,z]
[x + y + z + 1,
 x^2*z + y^2*x + z^2*y + 1,
 x^2*y + y^2*z + z^2*x + 1,
 x*y + x*z + y*z + x + y + z]

\end{verbatim} \normalsize
The Newton polytope of this threefold has the f-vector $(8, 16, 14, 6)$, and it
contains precisely $619$ lattice points.
 The eight vertices among them are

\small \begin{verbatim}

VERTICES
1 15 0 0 0
1  0 6 0 0
1  0 0 0 9
1  0 0 6 0
1  0 0 0 0
1 12 0 0 3
1  9 3 0 0
1  9 0 3 0

\end{verbatim} \normalsize
This four-dimensional polytope is the mixed fiber polytope (\ref{mfpimp})
for the tetrahedra $Q_1, Q_2, Q_3$ and the octahedron $Q_4$
specified in the file {\tt input}. \qed
\end{ex}

\smallskip

In (\ref{morphismG}) we assumed that the $g_i(t)$
are Laurent polynomials but this hypothesis can be
relaxed to other settings discussed in \cite{EKP, KE}.
In particular, the
Fundamental Theorem of Tropical Implicitization 
extends to the case when the $g_i(t)$ are rational functions.
Here is how this works in {\tt TrIm}.

\smallskip

\begin{ex} \rm
Let $\alpha_1, \ldots, \alpha_5$ and
$\beta_1, \ldots ,\beta_5$ be general complex numbers and 
consider the plane curve
which has the rational parametrization
\begin{equation}
\label{ratpara}
x \,\,= \,\, \frac{\alpha_1 t^3 +\alpha_2 t + \alpha_3}{\alpha_4 t^2 + \alpha_5}
\quad \hbox{and} \quad
y \,\,= \,\, \frac{\beta_1 t^4 + \beta_2 t^3 + \beta_3}{ \beta_4 t^2 + \beta_5}.
\end{equation}
This curve appears in \cite[Example 4.7]{EKP}. The {\tt input}
for {\tt TrIm} is as follows:

\small \begin{verbatim}

[ t, x, y ]
[ t^3 + t + 1   +  x*t^2 + x ,
  t^4 + t^3 + 1  +  y*t^2 + y ]
 
\end{verbatim} \normalsize

The equation of the plane curve is gotten by
eliminating the unknown $t$ from the two equations (\ref{ratpara}).
Tropical elimination using {\tt TrIm} predicts that the Newton polygon of that plane curve
is the following pentagon:
\small \begin{verbatim}

POINTS
1 4 2 
1 0 3 
1 2 3 
1 0 0 
1 4 0 

\end{verbatim} \normalsize
This prediction is the correct Newton polygon for generic
coefficients $\alpha_i$ and $\beta_j$.  In particular, generically the pairs of coefficients $(\alpha_4, \alpha_5)$ and $(\beta_4, \beta_5)$ in the denominators are distinct.  If we assume, however, that the denominators are the same, then the correct Newton polytope is a quadrangle inside this pentagon, as computed in \cite[Example 5.6]{EKP}.
 \qed
\end{ex}

\smallskip

\section{Tropical varieties}

We now explain the mathematics on which {\tt TrIm} is based.
The key idea is to embed the study of Newton polytopes
into the context  of tropical geometry
\cite{BJSST, DFS, ST, STY}. Let $I$ be any ideal in
the Laurent polynomial ring $\CC[x_1^{\pm 1}, \ldots, x_p^{\pm 1}]$.
Then its \emph{tropical variety} is
$$\T(I) \,\,\,= \,\,\,\bigl\{ w \in \RR^p \,:\, \text{in}_w(I) \text{ does not contain a monomial} \bigr\}.$$
Here ${\rm in}_w(I)$ is the ideal of $\CC[x_1^{\pm 1}, \ldots, x_p^{\pm 1}]$
which is generated by the
$w$-initial forms of all elements in $I$.
The set $\T(I)$ can be given the structure of a polyhedral fan, for instance,
by restricting the Gr\"obner fan of any homogenization of $I$.
 A point $w$ in $\cT(I)$ is called {\em regular} if it lies in the interior of a maximal cone in some fan structure on $\cT(I)$.  Every regular point $w$ naturally  
comes with a multiplicity $m_w$, which is a positive integer.  
We can define $m_w$ as the
sum of multiplicities of all minimal associate primes
of the initial ideal $\text{in}_w (I)$. The multiplicities 
$m_w$ on $\cT(I)$ are independent of the fan structure and they
satisfy the {\em balancing condition} \cite[Def.~3.3]{ST}.

If $I$ is a principal ideal, generated by one Laurent polynomial
$f(x)$, then $\cT(I)$ is the union of all codimension one cones
in the normal fan of the Newton polytope $P$ of $f(x)$. A point
$w \in \cT(I)$ is regular if and only if $w$ supports an edge of $P$,
and $m_w$ is the lattice length of that edge. It is important to
note that the polytope $P$ can be reconstructed uniquely,
up to translation, from the tropical hypersurface $\cT(I)$
together with its multiplicities $m_w$. The following
{\tt TrIm} example shows how to go back and forth between
the Newton polytope $P$ and its
tropical hypersurface $\cT(I)$.

\smallskip

\begin{ex} \rm \label{octahedron}
We  write the following polynomial onto the file {\tt poly}:

\small \begin{verbatim}

[ x, y, z ]
[ x + y + z + x^2*y^2 + x^2*z^2 + y^2*z^2 ]

\end{verbatim} \normalsize
The command {\tt ./TrCI.prl poly > fan} writes the 
tropical surface defined by this polynomial onto a file {\tt fan}.
That output file starts out like this:

\small \begin{verbatim}

AMBIENT_DIM
3

DIM
2
..........

\end{verbatim} \normalsize 
The tropical surface consists of 
$12$ two-dimensional cones  on $8$ rays in $\RR^3$. Three of the $12$ cones have multiplicity
two, while the others have multiplicity one. 
Combinatorially, this surface is the edge graph of the $3$-cube.
The Newton polytope $P$ is an octahedron, and it can be recovered from the data on
${\tt fan}$ after we place the $3 \times 3$ identity matrix in the file {\tt A.matrix}:

\small \begin{verbatim}

LINEAR_MAP
1 0 0 
0 1 0 
0 0 1

\end{verbatim} \normalsize
Our familiar command

\small \begin{verbatim}

./project.prl fan A.matrix 

\end{verbatim} \normalsize
now reproduces the Newton octahedron $P$ in the familiar format:

\small \begin{verbatim}

VERTICES
1 2 2 0 
1 0 2 2 
1 0 1 0 
1 2 0 2 
1 1 0 0 
1 0 0 1 

\end{verbatim} \normalsize
Note that the edge lengths of $P$
are the multiplicities on the
tropical surface. \qed
\end{ex}

The implementation of the command {\tt project.prl} is
based on the formula given in \cite[Theorem 5.2]{STY}.
See \cite[\S 2]{DFS} for a more general version. This result translates into an
algorithm for {\tt TrIm} which can be described as follows.
Given a generic vector $w \in \RR^k$ such that $\text{face}_w(P)$ is a vertex $v$, 
the $i^{th}$ coordinate $v_i$ is the number of intersections, counted with multiplicities, 
of the ray $w + \RR_{>0} e_i$ with the tropical variety.  Here, the multiplicity of the intersection with a cone $\Gamma$ is the multiplicity of $\Gamma$ times the absolute value of the $i^{th}$ coordinate 
of the primitive normal vector to the cone $\Gamma$. 

Intuitively, what we are doing in our software is the following.  We wish to determine the coordinates of the extreme vertex $v = \text{face}_w(P)$ of a polytope $P$ in a given direction $w$.  Our polytope is placed so that it lies in the positive orthant and touches all the coordinate hyperplanes.  To compute the $i^{th}$ coordinate of the vertex $v$, we can walk  from $v$ toward the $i^{th}$ hyperplane along the edges of $P$, while keeping track of the edge lengths in the $i^{th}$ direction. A systematic way to 
carry out the walk is to follow the edges whose inner normal cone intersect the ray $w + \RR_{>0} e_i$.  Recall that the multiplicity of a codimension one normal cone is the lattice length of the corresponding edge.  Using this subroutine for computing extreme vertices, the whole polytope is now
constructed using the method of Huggins \cite{iB4e}.

\smallskip

To compute the tropical variety $\cT(I)$ for an arbitrary ideal $I$
one can use the Gr\"obner-based software {\tt GFan} due to Jensen
\cite{BJSST, gfan}. Our polyhedral software {\tt TrIm} performs
the same computation faster when the generators of $I$
are Laurent polynomials $f_1,f_2,\ldots,f_c$
that are generic relative to their Newton polytopes
$P_1,P_2,\ldots,P_c$. It implements the following combinatorial formula
for the tropical variety $\cT(I)$ of the complete intersection~$I$.

\smallskip

\begin{theorem} \label{mixedmixed}
The tropical variety $\cT(I)$ is supported on a subfan of the
normal fan of the Minkowski sum $\sum_{i=1}^c P_i $.
A point $w $ is in $\cT(I)$ if and only if
the polytope ${\rm face}_w( \sum_{j \in J} P_j)$
 has dimension $\geq | J |$ for  $J \subseteq \{1,\ldots,c\}$.
The multiplicity of $\cT(I)$ at a regular point $w$ is
 the mixed volume
  \begin{equation}
\label{mixedvol} m_w \,\,\, = \,\,\, {\rm mixed}\,{\rm volume} \bigl( {\rm face}_w(P_1), {\rm face}_w(P_2), \ldots,
{\rm face}_w(P_c) \bigr),
\end{equation}
where we 
normalize volume respect to the 
affine lattice parallel to $\sum_{j \in J} P_j$.
\end{theorem}

\smallskip

For a proof of this theorem see \cite[\S 4]{ST}.
We already saw some examples in the second half of Section 2.
Here is one more such illustration:

\smallskip

\begin{ex} \rm
We consider the generic complete intersection of codimension
three in six-dimensional space $(\CC^*)^3$ given by the polynomials

\small \begin{verbatim}

[a,b,c,d,e,f]
[a*b*c*d*e*f + a + b + c + d + e + f + 1,
 a*b + b*c + c*d + d*e + e*f + f*a,
 a + b + c + d + e + f + 1]

\end{verbatim} \normalsize
The application of {\tt ./TrCI.prl} to this input internally
constructs the corresponding six-dimensional polytope
$P_1 + P_2 + P_3$. It lists
all facets and their normal vectors, it generates all 
three-dimensional cones in the normal fan, and it
picks a representative
vector $w$ in the relative interior of each such cone. The three polytopes
$\,{\rm face}_w(P_1)$, ${\rm face}_w(P_2)$ and ${\rm face}_w(P_3)\,$
are translated to lie in the same three-dimensional space,
and their mixed volume $m_w$ is computed.
If $m_w$ is positive then {\tt TrIm} outputs
the rays of that cone and the mixed volume $m_w$.
The output reveals that this tropical threefold in $\RR^6$ consists of  
$117$ three-dimensional cones  on $22$ rays. \qed
\end{ex}

\smallskip

In our implementation of Theorem \ref{mixedmixed}
in {\tt TrIm}, the Minkowski sum is computed using the {\tt iB4e} 
library \cite{iB4e}, enumerating the $d$-dimensional cones is 
done by {\tt Polymake} \cite{GJ}, the mixed volumes are computed 
using the {\tt mixed volume library} \cite{EC}, and integer 
linear algebra for lattice indices is done using {\tt NTL} \cite{ntl}.   
 What oversees all the computations is the {\tt perl} script {\tt TrCI.prl}.  
The output format is consistent with that of the current 
version of {\tt Gfan} \cite{gfan} and is intended to interface 
with the polyhedral software {\tt Polymake} \cite{GJ} when it 
supports polyhedral complexes and fans in the future.  

\smallskip

Elimination theory is concerned with 
computing the image of an algebraic variety
whose ideal $I$ we know under
a morphism $\alpha : (\CC^*)^p \rightarrow (\CC^*)^r$. 
We write $\beta$ for the corresponding homomorphism
from the Laurent polynomial ring in $r$ unknowns to the
Laurent polynomial ring in $p$ unknowns. Then
$J = \beta^{-1}(I)$ is the ideal of the image variety,
and, ideally, we would like to find generators for $J$.
That problem is  too hard, and what we do instead
is to apply tropical elimination theory as follows.
We assume that $\alpha$ is a monomial map,
specified by an $r \times p$ integer matrix ${\bf A} = (a_{ij})$ 
as in (\ref{monomap}). Rather than computing the variety 
of $J$ from the variety of $I$, we instead compute the
tropical variety $\cT(J)$ from the tropical variety $\cT(I)$.
This is done by the following theorem which characterizes
the multiplicities.

\smallskip

\begin{theorem} 
\label{superdupertheorem}
The tropical variety $\cT(J)$ equals the image of
$\cT(I)$ under the linear map $\mbA$.
If the monomial map $\alpha$ induces
a generically finite morphism of degree $\delta$ from
the variety of $I$ onto the variety of $J$ then the
multiplicity of $\cT(J)$ at a regular point $w$ is computed by the formula
\begin{equation}\label{superduperformula}
m_w \,\,= \,\, \frac{1}{\delta}\cdot \sum_{v}  m_v \cdot
 \ind(\bL_w \cap\ZZ^r : \mbA(\bL_v\cap\ZZ^p)).
\end{equation}
The sum is over all points $v$ in $\cT(I)$ with
$\mbA v = w$. We assume that the number
of these points is finite. they are all regular in $\cT(I)$,
and $\bL_v$ is the linear span of a neighborhood of $v$ in
$\cT(I)$, and similarly for $w \in \cT(J)$.
\end{theorem}

\smallskip

The formula (\ref{superduperformula}) can be regarded as a 
push-forward formula in intersection theory on toric varieties,
and it constitutes the workhorse inside 
the {\tt TrIm} command {\tt project.prl}.  
When the tropical variety $\cT(J)$ has codimension
one, then that tropical hypersurface determines the Newton polytope of
the generator of $J$.
The transformation from tropical hypersurface
to mixed fiber polytope was behind all our earlier examples.
That transformation was shown explicitly for an octahedron
in Example \ref{octahedron}.

In all our examples so far, the ideal $I$ was
tacitly assumed to be a generic complete intersection,
and Theorem \ref{mixedmixed} was used to
determine the tropical variety $\cT(I)$ and 
the multiplicities $m_v$.  In other words, the
command {\tt TrCI} furnished the ingredients
for the formula (\ref{superduperformula}).
In particular, then $I$ is the ideal of the graph of a morphism,
as in Section 5, then Theorem \ref{superdupertheorem} specializes
to the formula for tropical implicitization given in \cite{STY}.

It is important to note, however, that Theorem \ref{superdupertheorem}
applies to any ideal $I$ whose tropical variety happens to be known,
even if $I$ is not a generic complete intersection. For instance,
$\cT(I)$ might be the output of a {\tt GFan} computation, or it
might be one of the special tropical varieties which have
already been described in the literature. Any tropical variety with
known multiplicities can serve as the input to the {\tt TrIm} command {\tt project.prl}.

\bigskip

{\bf Acknowledgments. }
We are grateful to Peter Huggins for his help
at many stages of the {\tt TrIm} project. His {\tt iB4e} software
became a crucial ingredient in our implementation.
Bernd Sturmfels was partially supported by the National 
Science Foundation (DMS-0456960), and Josephine Yu was supported
by a UC Berkeley Graduate Opportunity Fellowship and by the 
Institute for Mathematics and its Applications (IMA) in Minneapolis.

 \end{document}